\documentclass[prb,superscriptaddress,twocolumn,floatfix,amsmath,amssymb,showpacs]{revtex4}

\usepackage{graphicx}
\usepackage{dcolumn}
\usepackage{verbatim}
\usepackage{times}
\usepackage{subfigure}
\usepackage{bm}
\usepackage{color}
\usepackage{subfigure}
\usepackage[colorlinks,bookmarks=false,citecolor=blue,linkcolor=red,urlcolor=blue,backref=red,dvipdfm]{hyperref}
\usepackage{booktabs}
\usepackage{appendix}

\begin{document}

\title{Magnetic and topological transitions in three-dimensional topological Kondo insulator\footnote{Supported by the National Natural
Science Foundation of China under Grant Nos 11764010, 11504061, 11564008, 11704084, 11704166, Guangxi NSF under Grant Nos 2017GXNSFAA198169, 2017GXNSFBA198115, and SPC-Lab Research Fund (No. XKFZ201605).}}

\author{Huan Li\footnote{Corresponding author, Email: lihuan@glut.edu.cn}}
\affiliation{College of Science, Guilin University of Technology, Guilin 541004, China}

\author{Zhi-Yong Wang}
\affiliation{College of Science, Guilin University of Technology, Guilin 541004, China}

\author{Xiao-Jun Zheng\footnote{Corresponding author, Email: zhengxiaojun323@foxmail.com}}
\affiliation{College of Science, Guilin University of Technology, Guilin 541004, China}

\author{Yu Liu}
\affiliation{Institute of Applied Physics and Computational Mathematics, Beijing 100088, China}
\affiliation{Software Center for High Performance Numerical Simulation, China Academy of Engineering Physics, Beijing 100088, China}

\author{Yin Zhong}
\affiliation{Center for Interdisciplinary Studies $\&$ Key Laboratory for Magnetism and Magnetic Materials of the MoE, Lanzhou University, Lanzhou 730000, China}

\begin{abstract}

By using an extended slave-boson method, we draw a global phase diagram summarizing both magnetic phases and paramagnetic (PM) topological insulating phases (TI$_s$) in three-dimensional topological Kondo insulator (TKI). By including electron hopping (EH) up to third neighbor, we identify four strong topological insulating (STI) phases and two weak topological insulating (WTI) phases, then the PM phase diagrams characterizing topological transitions between these TI$_s$ are depicted as functions of EH, $f$-electron energy level and hybridization constant. We also find an insulator-metal transition from a STI phase which has surface Fermi rings and spin textures in qualitative agreement to TKI candidate SmB$_6$. In weak hybridization regime, antiferromagnetic (AF) order naturally arises in the phase diagrams, and depending on how the magnetic boundary crosses the PM topological transition lines, AF phases are classified into AF topological insulator (AFTI) and non-topological AF insulator (nAFI), according to their $\mathcal{Z}_2$ indices. In two small regions of parameter space, two distinct topological transition processes between AF phases occur, leading to two types of AFTI, showing distinguishable surface dispersions around their Dirac points.
\end{abstract}



\maketitle

Over the recent years, searching topological phases of matter has becoming one of the central topics in condensed matter physics.~\cite{Zhang11} Among the enlarging family of topological matters, the strongly correlated electron systems offer as important basis, because they naturally involve rich kinds of mechanism,
hence can generate a variety of interacting topological phases, such as interacting topological insulator,~\cite{Go12} topological Mott insulator,~\cite{Yu11} interacting topological superconductor,~\cite{Wang12} Weyl semimetal,~\cite{Wan11} topological Kondo insulator,~\cite{Dzero12} antiferromagnetic topological insulator (AFTI),~\cite{Mong10,Fang13,ZhiLi15,Li18} etc.

Topological Kondo insulator (TKI),~\cite{Dzero12} a heavy-fermion system with strong Coulomb interaction and $d$-$f$ hybridization governing by spin-orbit coupling, preserves time-reversal symmetry (TRS), therefore can generate topological insulating phases (TI$_s$) with Kondo screening effect. As revealed by previous works, variation of electron hopping (EH) strength, $f$-electron energy level $\epsilon_f$ and hybridization constant $V$ can drive topological transitions among phases of strong topological insulator (STI), weak topological insulator (WTI) and normal Kondo insulator (nKI).~\cite{Legner14} However, existing works in literature are restricted to their interested parameter regime, hence the studied TI$_s$ are still confined to a limited number of STI and WTI, and the full STI and WTI phases in TKI have not been explored adequately, particularly at the presence of strong electron-electron correlation.~\cite{Legner14} In this work, by considering adequate parameter space of periodic Anderson model (PAM), we uncover all possible TI$_s$ in three-dimensional (3D) TKI: four STI and two WTI, each possessing distinct surface states and Dirac cones. We also present the paramagnetic (PM) phase diagrams characterizing topological transitions between these TI$_s$, as functions of EH, $\epsilon_f$ and $V$. By proper fitting of EH, we verify a STI phase with Fermi surfaces and spin textures which can qualitatively simulate the TKI material SmB$_6$,~\cite{Xu16} confirming the applicability of PAM to TKI, and it is find that this STI phase is in vicinity to an insulator-metal transition driving by enhancement of $V$.

In heavy-fermion systems, the interplay and competition between Kondo screening and magnetic correlation can motivate magnetic transitions when the Kondo interaction is reduced.~\cite{Vekic95} Similarly, in half-filled TKI, theoretical calculations have verified a transition to AF phase when the hybridization interaction $V$ is weakened,~\cite{Li18,Peters18} reminiscent of the induced magnetism in pressurized SmB$_6$.~\cite{Butch16,Zhao17,Chang17} Besides,
our earlier work has proved that due to the combined $\mathcal{S}$ symmetry of time reversal and translation operations, the AF states in TKI remain topological distinguishable, regardless of the breaking of TRS by magnetic order. We has developed a $\mathcal{Z}_2$ topological classification to the AF states in TKI and proposed a novel AFTI phase under unique setting of model parameters, together with an AFTI-nontopological AF insulator (nAFI) topological transition while EH was shifted in some way.~\cite{Li18} Unfortunately, why AFTI should appear in such parameter region is still not clear, and it remains confused whether new AF phases exist in other parameter regions.
We have shown that at least near the magnetic boundary (MB), the $\mathcal{Z}_2$ index for AF directly relies on that of TI phase from which the AF order develops,~\cite{Li18} therefore, in order to investigate all possible AF phases with distinct topologies, the magnetic transition and classification of AF phases should be discussed on the basis of the PM phase diagrams summarizing all TI$_s$, i.e., the four STI and two WTI should be included properly to study the AF transition as well as the topological transitions between AF phases.

We use the spin-1/2 PAM to character the 3D TKI in cubic lattice:~\cite{Alexandrov15}
\begin{align}
H=H_{d}+H_{f}+H_{df}+H_{U},
\label{PAM}\end{align}
where $H_{d}=\sum_{\mathbf{k},\alpha}(\epsilon^d_\mathbf{k}-\mu)d^\dag_{\mathbf{k}\alpha}d_{\mathbf{k}\alpha}$,
$H_{f}=\sum_{\mathbf{k},\alpha}(\epsilon_f+\epsilon^f_\mathbf{k}-\mu)f^\dag_{\mathbf{k}\alpha}f_{\mathbf{k}\alpha}$.
$
H_{df}=V\sum_{\mathbf{k},\alpha,\beta}\mathbf{S}_{\mathbf{k}}\cdot \vec{\sigma}_{\alpha\beta}d^\dag_{\mathbf{k}\alpha}f_{\mathbf{k}\beta}+h.c.
$
is the Kondo hybridization with spin-orbit coupling, in which $\mathbf{S}_\mathbf{k}=(\sin \mathbf{k}\cdot\mathbf{a}_1,\sin \mathbf{k}\cdot\mathbf{a}_2,\sin \mathbf{k}\cdot\mathbf{a}_3)$,~\cite{Alexandrov15} with the element vectors $\mathbf{a}_1$, $\mathbf{a}_2$, $\mathbf{a}_3$ for cubic lattice. $H_U=U\sum_in^f_{i\uparrow}n^f_{i\downarrow}$ is the on-site coulomb repulsion between $f$ electrons, and we consider infinite $U$ in this work. We includes EH up to third neighbor, with $t_{d(f)}$, $t^{\prime}_{d(f)}$, $t^{\prime\prime}_{d(f)}$ denote nearest-neighbor (NN), next-nearest-neighbor (NNN), and next-next-nearest-neighbor (NNNN) hopping amplitudes, respectively, which determine the tight-binding dispersions $\epsilon^{d(f)}_\mathbf{k}$. The chemical potential $\mu$ is used to fix the total electron number to half filling $n_t=2$, and variable EH, hybridization interaction $V$ and $f$ energy level $\epsilon_f$ are considered. In what follows, $t_d=1$ is set as energy unit, and we choose $t_f=-0.2$ and keep $t^\prime_d/t_d=t^\prime_f/t_f$ to get a medium gapped insulating phase (unless when the insulator-metal transition is discussed).

We employ the Kotliar-Ruckenstein (K-R) slave-boson method~\cite{Kotliar86,Sun93,Li18} to solve PAM. Similar to Coleman's slave boson theory,~\cite{Alexandrov15} the mean-field approximation of PAM Eq. \ref{PAM} in large-$U$ limit reads~\cite{Li18}
\begin{align}
&H_{MF}=N(-\eta n_f)\nonumber\\
&+\sum_{\mathbf{k},\alpha,\beta}(d^\dag_{\mathbf{k}\alpha},f^\dag_{\mathbf{k}\alpha})
\left(
\begin{array}{cc}
(\epsilon^d_\mathbf{k}-\mu)\delta_{\alpha\beta}&\tilde{V}\mathbf{S}_{\mathbf{k}}\cdot \vec{\sigma}_{\alpha\beta}\\
\tilde{V}\mathbf{S}_{\mathbf{k}}\cdot \vec{\sigma}_{\alpha\beta}&(\tilde{\epsilon}^f_\mathbf{k}-\mu)\delta_{\alpha\beta}
\end{array}
\right)
\left(\begin{array}{cc}
d^\dag_{\mathbf{k}\beta}\\f^\dag_{\mathbf{k}\beta}
\end{array}
\right),
\label{PM}\end{align}
where the effective hybridization $\tilde{V}=VZ$ is renormalized by factor $Z=\sqrt{2(1-n_f)/(2-n_f)}$, and the effective $f$ dispersion $\tilde{\epsilon}^f_{\mathbf{k}}=\epsilon_f+\eta+Z^2\epsilon^f_{\mathbf{k}}$, in which $\eta$ shifts the $f$ level. $n_f$ is the density of $f$ electron per site, N is the number of lattice sites. The PM mean field parameters $n_f$, $\eta$, $\mu$ are solvable through saddle-point solution for $H_{MF}$, then the quasi-particle dispersions which are the eigenvalues of the Hamiltonian matrix in Eq. \ref{PM} (in a modified form) are used to identify the $\mathcal{Z}_2$ index.

\heavyrulewidth=1bp

\begin{table*}
\small
\renewcommand\arraystretch{0.6}
\caption{\label{table1}
Parameters and $\mathcal{Z}_2$ invariants of the TI phases shown in Fig. \ref{fig1}. In all phases, $t_f=-0.2$.}
\begin{tabular*}{17cm}{@{\extracolsep{\fill}}cccccccccccccc}
\toprule
    phase & $t^\prime_d$  & $t^{\prime\prime}_d$ &  $t^\prime_f$ & $t^{\prime\prime}_f$ & $\epsilon_f$ & $V$ & $\delta_\Gamma$ & $\delta_X$ & $\delta_M$ & $\delta_R$ & $\nu_0$ & $\nu_j$ & Dirac points\footnotemark[1]\\
\hline

 STI$_{\bar{\Gamma}}$ & 0.26 & 0.26 & -0.052 & -0.052 & -2 & 0.7 & 1 & -1 & -1 & -1 & 1 & - &$\bar{\Gamma}$\\
STI$_{\bar{\Gamma}\bar{X}}$ & -0.35 & -0.35 & 0.07 & 0.07 & -2 & 1 & -1 & 1 & -1 & -1 & 1 & - &$\bar{\Gamma}$, $\bar{X}$\\
STI$_{\bar{M}}$ & 0.252 &0.252 & -0.0504 & -0.0504 & -2 & 1.5 & 1 & 1 & 1 & -1 & 1 & - & $\bar{M}$\\
STI$_{\bar{M}\bar{X}}$ & 0.4 & 0 & -0.08 & 0 & -1 & 1   & 1 & 1 & -1 & 1 & 1 & - &$\bar{M}$, $\bar{X}$\\
WTI$_{\bar{\Gamma}\bar{M}}$ & -0.6 & 0 & 0.12 & 0 & -1 & 1  & -1 & 1 & 1 & -1 & 0 & 1 & $\bar{\Gamma}$, $\bar{M}$\\
WTI$_{\bar{X}}$ & -0.2 & 0 & 0.04 & 0 & -2 & 1     & 1 & 1 & -1 & -1 & 0 & 1 &$\bar{X}$\\

\bottomrule
\end{tabular*}
\footnotetext[1]{The surface dispersions are calculated on (001) surface.}

\label{data}
\end{table*}


\begin{figure}[tbp]
\hspace{-0.3cm} \includegraphics[totalheight=3.6in]{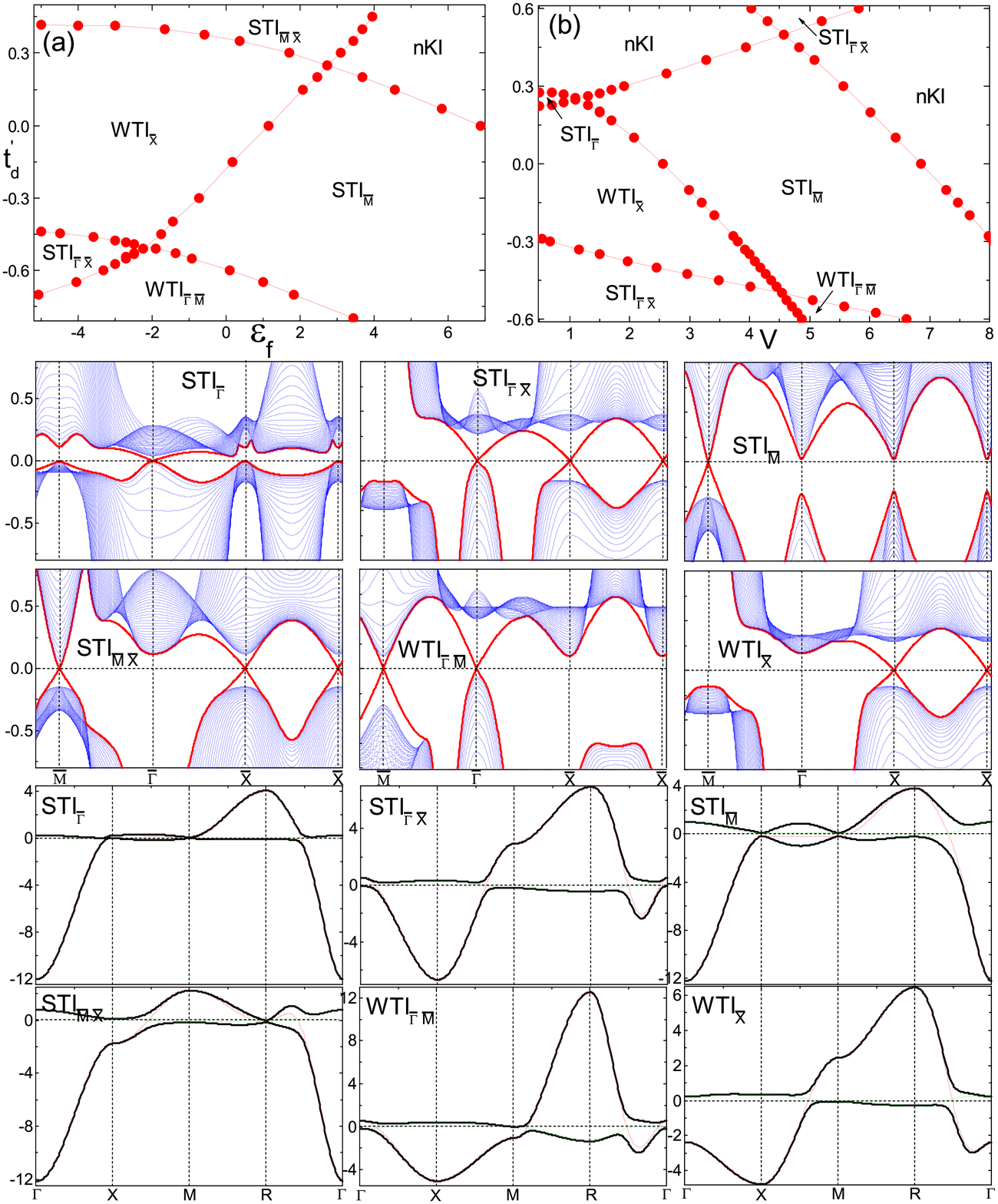}
\caption{First row: PM phase diagrams of 3D TKI. Parameters: $t_f=-0.2$, $t^\prime_f/t_f=t^\prime_d/t_d$, (a) $V=1$, $t^{\prime\prime}_d=t^{\prime\prime}_f=0$; (b) $\epsilon_f=-2$, $t^{\prime\prime}_d=t^{\prime}_d$, $t^{\prime\prime}_f=t^{\prime}_f$. Second and third rows: slab dispersions of TI$_s$. The red lines denote the surface states. Last two rows: quasi-particle dispersions (black solid lines) in six TI$_s$, the red and green lines are $d$- and renormalized $f$- dispersions, respectively. Parameters and $\mathcal{Z}_2$ invariants for each TI$_s$ are listed in Tab. \ref{table1}.
}
\label{fig1}
\end{figure}

In last two rows of Fig. \ref{fig1}, we show six types of distinct quasi-particle spectrums of PM TI$_s$, each with different model parameters listed in Tab. \ref{table1}, comparing with $d$ and $f$ dispersions. At the eight high symmetric points (HSP$_s$) $\mathbf{k}_m$ in 3D Brillouin zone (BZ) (i.e.,$\Gamma$=(0,0,0); $X$=($\pi$,0,0), (0,$\pi$,0), (0,0,$\pi$); $M$=($\pi$,$\pi$,0), ($\pi$,0,$\pi$), (0,$\pi$,$\pi$); and $R$=($\pi$,$\pi$,$\pi$)), the hybridization vanishes (due to its odd parity), consequently the quasi-particle energy equals either $\epsilon^d_{\mathbf{k}}$ or $\tilde{\epsilon}^f_{\mathbf{k}}$, leading to the parity of occupied states $\delta_m=$ 1 or -1 at $\mathbf{k}_m$, respectively. Therefore, the strong topological index $\nu_0$ can be easily obtained by observing the bulk dispersions in Fig. \ref{fig1} via $(-1)^{\nu_0}=\prod_{m\in \mathrm{HSP}_s} \delta_m$, and the weak topological indices $\nu_j$ ($j=x,y,z$) are calculated from the HSP$_s$ on $k_j$ plane $P_j$ through $(-1)^{\nu_j}=\prod_{m\in P_j} \delta_m$.~\cite{Dzero12} The quantities $\delta_m$ and $\mathcal{Z}_2$ indices for six TI$_s$ are listed in Tab. \ref{table1}.

By diagonalizing 40 slabs to simulate the 3D lattice with opened (001) surface, the surface states of the six TI$_s$ are computed and displayed by second and third rows in Fig. \ref{fig1}. On (001) surface, there are four HSP$_s$: $\bar{\Gamma}$=(0,0); $\bar{X}$=($\pi$,0), (0,$\pi$); and $\bar{M}$=($\pi$,$\pi$), each of the six TI$_s$ in Fig. \ref{fig1} has Dirac points locating at different HSP$_s$. The requirement of odd number of Dirac points on surface of STI leads to four inequivalent STI$_s$: STI$_{\bar{\Gamma}}$, STI$_{\bar{M}}$, STI$_{\bar{\Gamma}\bar{X}}$, and STI$_{\bar{M}\bar{X}}$, in which the
subscripts denote the locations of Dirac points. For WTI$_s$, there are even number of Dirac points, resulting in two WTI$_s$: WTI$_{\bar{\Gamma}\bar{M}}$ and WTI$_{\bar{X}}$. For nKI, generally no Dirac point exists, however, there is a special nKI with Dirac points at all four HSP$_s$, since Fermi level crosses its surface states even times between two arbitrary HSP$_s$, this phase is actually non-topological phase rather than a topological one.

\begin{figure}[tbp]
\hspace{-0.1cm} \includegraphics[totalheight=2.52in]{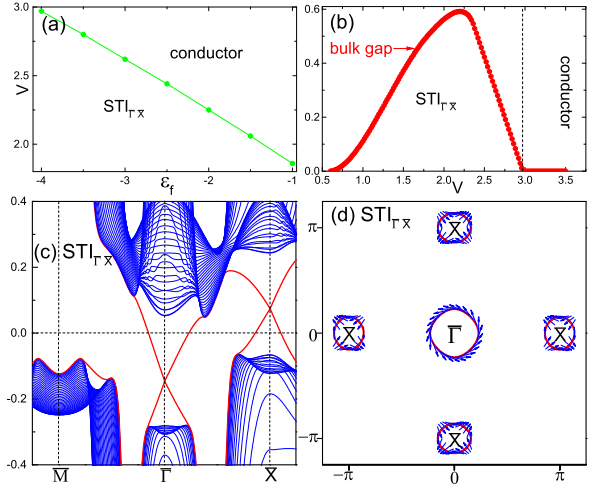}
\caption{(a) Insulator-metal transition from STI$_{\bar{\Gamma}\bar{X}}$ to conducting phase. (b) Closing of bulk gap during this transition. (c) Slab dispersions of STI$_{\bar{\Gamma}\bar{X}}$. (d) Surface Fermi rings and spin textures of STI$_{\bar{\Gamma}\bar{X}}$. $t^\prime_d=t^{\prime\prime}_d=-0.375$, $t_f=-0.23$, $t^\prime_f=t^{\prime\prime}_f=0.2$. $\epsilon_f=-4$ for (b) to (d), and $V=1$ for (c) and (d).
}
\label{fig2}
\end{figure}

In the first row of Fig. \ref{fig1}, with varying $t^\prime_d$, $\epsilon_f$ and $V$, we have located the topological boundaries among all possible TI$_s$, determining by the change of $\mathcal{Z}_2$ index. The topological transitions between TI$_s$ are generated by closing and reopening of the insulating gap at certain HSP, leading to an inversion of parity and consequently the shifting of $\mathcal{Z}_2$ index.~\cite{Li18}


\begin{figure}[tbp]
\hspace{-0.1cm} \includegraphics[totalheight=2.25in]{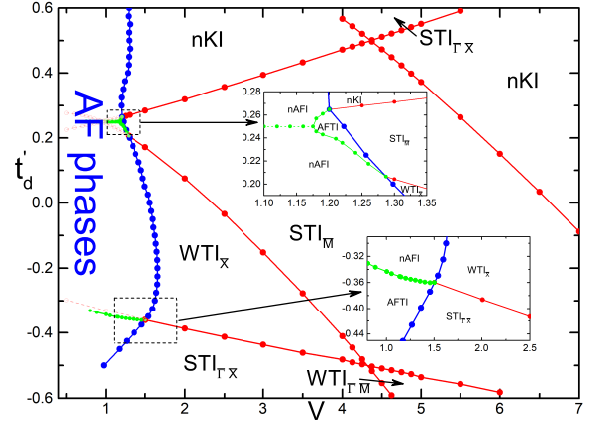}
\caption{(a) Magnetic boundary in 3D TKI (blue lines). Near $t^\prime_d=0.25$ and $-0.35$, topological transitions between AF phases take place (green solid lines). Parameters: $\epsilon_f=-1.5$, $t^{\prime\prime}_d=t^{\prime}_d$, $t^{\prime\prime}_f=t^{\prime}_f$, $t_f=-0.2$ and $t^\prime_f/t_f=t^\prime_d/t_d$.
}
\label{fig3}
\end{figure}

In above, we have set $t^\prime_f/t_f=t^\prime_d/t_d$, under which the Dirac points in TI$_s$ all cross the Fermi energy, leading to the vanishment of Fermi surface. For TKI candidate SmB$_6$, medium-sized surface Fermi rings around $\bar{\Gamma}$ and $\bar{X}$ were detected through ARPES, verifying it in a STI$_{\bar{\Gamma}\bar{X}}$ phase.~\cite{Xu16} Though our study of TKI is based on the simplified PAM, it can still produce a STI$_{\bar{\Gamma}\bar{X}}$ with similar surface states to SmB$_6$. To do this, we chose EH departing from $t^\prime_f/t_f=t^\prime_d/t_d$, and found a STI$_{\bar{\Gamma}\bar{X}}$ phase with Fermi surfaces and helical spin textures quite similar to SmB$_6$, see Fig. \ref{fig2}(c) and (d). Furthermore, this phase is in vicinity to an insulator-metal transition generated by shifting of $V$ or $\epsilon_f$ (see Fig. \ref{fig2}(a) and (b)), which may be account for the metallic phase observed in pressurized SmB$_6$.~\cite{Paraskevas15,Zhao17}

Based on the PM phase diagrams of TI$_s$, we now study the AF transitions in TKI. In our previous work, the original K-R method of symmetric PAM~\cite{Kotliar86,Sun93} has been generalized to treat AF phases in non-symmetric case,~\cite{Li18} which can be applied to TKI. The resulting mean-field Hamiltonian is rather complicated in that in addition to $n_f$, $\eta$ and $\mu$, two AF order parameters $m_f$ and $h$ should be determined, besides, two renormalization factors $Z_1$ and $Z_2$ arise. Due to the $\mathcal{S}$ symmetry combined by TRS and lattice translation, the AF phases in TKI fall into $\mathcal{Z}_2$ topological class, and the $\mathcal{Z}_2$ index $\nu$ is calculated from the parities of the occupied spectrums at four Kramers degenerate momenta (KDM) $\mathbf{p}_m$ ($\Gamma$ and three $X$ points) via $(-1)^{\nu}=\prod_{\mathbf{p}_m\in \mathrm{KDM}}\delta_m$, in which $\delta_m=\prod_i\xi_i(\mathbf{p}_m)$, with the parity $\xi_i(\mathbf{p}_m)$ of $i$-th occupied state at $\mathbf{p}_m$ equals either 1 or -1, when quasi-particle energy equals that of $d$ or $f$ at $\mathbf{p}_m$, respectively.~\cite{Li18} Particularly, the strong topological index $\nu_0$ on the PM side of the MB directly determines $\nu$ of the AF phase near MB, namely, $\nu_0=1$ (STI) leads to $\nu=1$ (AFTI), while $\nu_0=0$ (WTI or nKI) leads to $\nu=0$ (nAFI),~\cite{Mong10,Li18} giving a straightforward verification of the AF phases near MB.

\begin{figure}[tbp]
\hspace{-0.1cm} \includegraphics[totalheight=2.115in]{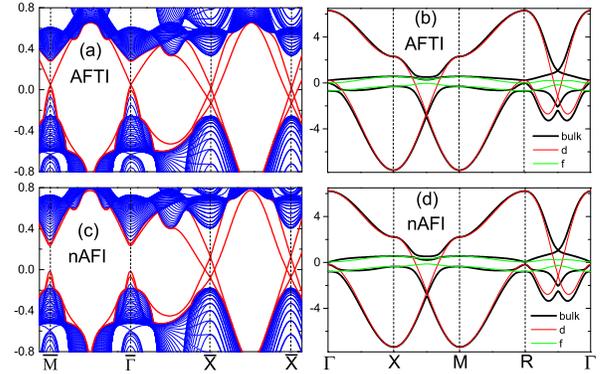}
\caption{Surface dispersions (left column) and quasi-particle dispersions (right column) of AFTI and nAFI near the AF topological boundary in lower inset of Fig. \ref{fig3}. $V=1.4$, $t^\prime_d=-0.37$ for AFTI, and $t^\prime_d=-0.35$ for nAFI.
}
\label{fig4}
\end{figure}

\begin{figure}[tbp]
\hspace{-0.1cm} \includegraphics[totalheight=2.79in]{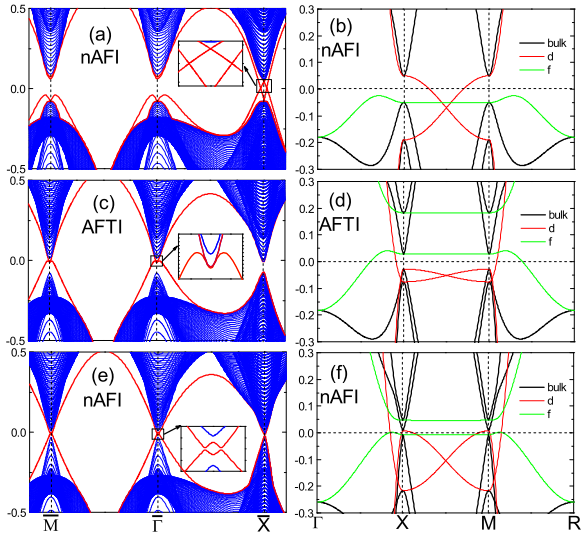}
\caption{Surface spectrums (left column) and quasi-particle dispersions (right column) of the three AF phases near the AF topological boundary in upper inset of Fig. \ref{fig3}. $V=1.19$ for all. From up, middle to down rows, $t^\prime_d=0.235$, $0.253$, and $0.264$, respectively.
}
\label{fig5}
\end{figure}

The magnetic critical hybridization $V_c$ is calculated as a function of $t^\prime_d$, then the MB plotted by $V_c$ is added to the phase diagram, see Fig. \ref{fig3}.
The MB crosses the topological boundaries of TI$_s$ in two parts, one near $t^\prime_d=0.25$, the other around $t^\prime_d=-0.35$, see insets of Fig. \ref{fig3} for details.

Around $t^\prime_d=-0.35$, the MB is divided by the STI$_{\bar{\Gamma}\bar{X}}$-WTI$_{\bar{X}}$ transition line into two parts, leading to AFTI and nAFI just below the two parts of MB, respectively. While $V$ is lowered further from MB, $\nu$ of AF phases should be computed from the 3D spectrums (e.g., Fig. \ref{fig4} (b) and (d)) to determine the AFTI-nAFI topological boundary, which is demonstrated by the green line near $t^\prime_d=-0.35$ in Fig. \ref{fig3}. The AFTI-nAFI transition is realized via parity inversion during gap closing and reopening at $\Gamma$, and it converges with STI$_{\bar{\Gamma}\bar{X}}$-WTI$_{\bar{X}}$ boundary at the MB, see the lower inset in Fig. \ref{fig3}.

Near $t^\prime_d=0.25$, the MB is separated by WTI$_{\bar{X}}$-STI$_{\bar{M}}$ and STI$_{\bar{M}}$-nKI lines into three parts. Below the middle part of MB (which touches STI$_{\bar{M}}$), an AFTI arises, while below the other two parts of MB, nAFI emerges. The AFTI-nAFI transition forms a narrow water-drop-shaped area in which AFTI survives (green solid line in the upper inset in Fig. \ref{fig3}). Besides, though nAFI$_s$ above and below $t^\prime_d=0.25$ have quite different dispersions (compare Fig. \ref{fig5}(b) with (f)), they still have equal $\nu=0$, since their magnetic orders grow from nKI and WTI, respectively. Though band gap is closed at the boundary between two nAFI$_s$, no parity inversion occurs, consequently no topological transition takes place (see the green dashed line in upper inset of Fig. \ref{fig3}).

The surface states of AF phases are shown in Fig. \ref{fig4}(a), (c) near $t^\prime_d=-0.35$, and in Fig. \ref{fig5}(a), (c), (e) around $t^\prime_d=0.25$, respectively. In AFTI$_s$, the Dirac points at $\bar{\Gamma}$ and $\bar{M}$ are protected by topology hence are robust, see Fig. \ref{fig4}(a) and Fig. \ref{fig5}(c). Furthermore, the Dirac surface states of two AFTI$_s$ (one near STI$_{\bar{M}}$ and the other near STI$_{\bar{\Gamma}\bar{X}}$) disperse quite differently, in which the former constructs a valley shape (Fig. \ref{fig5}(c)). In contrast, the gapless surface states at $\bar{X}$ in both AFTI and nAFI (Fig. \ref{fig4}(a), (c) and Fig. \ref{fig5}(a)) are not robust, since they can be gapped by additional factor such as gate voltage.~\cite{Li18}

In summary, we have performed a slave-boson mean-field analysis of the 3D TKI using spin-orbit coupled PAM, and presented the phase diagrams including all possible PM TI$_s$ in TKI: four STI and two WTI, each with distinguishable locations of Dirac points. We also obtained a STI$_{\bar{\Gamma}\bar{X}}$ phase with similar surface states to SmB$_6$, and found it can be driven to conducting state through an insulator-metal transition by enhanced hybridization. We also investigated the magnetic boundary of AF phases in TKI, and found the topological transitions between AFTI and nAFI in two narrow regions in parameter space. Besides, we found two types of AFTI with distinct dispersions at the Dirac points. Though our work is based on an uniform mean-field approximation, any site-dependent treatment will not break the application of $\mathcal{Z}_2$ classification of both PM and AF states.~\cite{Peters18,Chang17} We hope our work can help to reach a comprehensive understanding of novel AFTI phases in strongly correlated electrons systems.


\end{document}